# Puffing and micro-explosion behavior in combustion of butanol/jet A-1 and acetone-butanol-ethanol (A-B-E)/jet A-1 fuel droplets

D. Chaitanya Kumar Rao[a], Srinibas Karmakar[a,†], S. K. Som[b]

[a]*Department of Aerospace Engineering, Indian Institute of Technology Kharagpur, West Bengal, 721302, India*
[b]*Department of Mechanical Engineering, Indian Institute of Technology Kharagpur, West Bengal, 721302, India*

ABSTRACT

The present investigation deals with the puffing and micro-explosion characteristics in the combustion of a single droplet comprising butanol/Jet A-1, acetone-butanol-ethanol (A-B-E)/Jet A-1 blends, and A-B-E. The onset of nucleation, growth of vapor bubble and subsequent breakup of droplet for various fuel blends have been analyzed from the high-speed images. Puffing was observed to be the dominant phenomenon in 30% butanol blend, while micro-explosion was found to be the dominant one in other fuel blends (blend with 50% butanol or 30% A-B-E or 50% A-B-E). It was observed that puffing always preceded the micro-explosion. The probability of micro-explosion in droplets with A-B-E blends was found to be higher than that of butanol blends. Although the rate of bubble growth was almost similar for all butanol and A-B-E blends, the final bubble diameter before the droplet breakup was found to be higher for 50/50 blends than that of 30/70 blends. The occurrence of micro-explosion shortened the droplet lifetime, and this effect appeared to be stronger for droplets with 50/50 composition. Micro-explosion led to the ejection of both larger and smaller secondary droplets; however, puffing resulted in relatively smaller secondary droplets compared to micro-explosion. Puffing/micro-explosion were also observed in the secondary droplets.

Keywords: Jet A-1, butanol, acetone-butanol-ethanol, micro-explosion, puffing, bubble growth

## 1. Introduction

Alcohols, such as methanol, ethanol and butanol are considered as potential alternative fuels. Butanol, a longer carbon chain alcohol, is considered as a promising candidate due to its advantages over the short chain alcohols (ethanol and methanol) such as higher energy density and less hygroscopic in nature [1]. Studies on butanol-jet fuel blends are limited. Dziegielewski et al. [2]

†Corresponding author. Tel.: +91 3222 283012
E-mail address: skarmakar@aero.iitkgp.ernet.in (Srinibas Karmakar)

studied the compatibility of butanol and bio-butanol blending in Jet-A and diesel fuels, and it was suggested that 10% blending of butanol is acceptable in both Jet A-1 and diesel. They also reported that the blends of butanol reduced the flash point and significantly influenced the conductivity of Jet fuel. Mendez et al. [3] concluded that blends of butanol with jet fuel present promising performance. Alam et al. [4] reported an absence of soot shell and complete combustion of butanol droplets. Butanol is mainly produced by acetone–butanol–ethanol (A-B-E) fermentation that uses bacterial fermentation to produce acetone, n-butanol, and ethanol from biomass. The fermentation products contain acetone, butanol, and ethanol with a volumetric ratio of approximately 3:6:1. Despite the advantages of A-B-E fermentation, the higher costs of separation of butanol from the fermentation broth have prohibited the large-scale industrial production of butanol [5-7]. The direct use of the intermediate product (acetone-butanol-ethanol mixture) could be an economical pathway if used for clean combustion. Chang et al. [8] found that the ABE–diesel blends even with small amount of water were stable in the stability tests. Recently, the spray combustion characteristics of A-B-E and diesel blends were investigated in a constant volume chamber, and it was reported that ABE–diesel blends presented better combustion efficiency and lower soot emission [9-11]. Occurrences of Micro-explosions were also reported due to the higher volatility differential among the components of ABE-diesel blends, ultimately improving the combustion performance. Ma et al. [12] studied the evaporation characteristics of A-B-E and diesel blended droplets at high ambient temperatures and reported bubble formation and its rupture at high temperatures.

The fragmentation of droplets by puffing/micro-explosion plays a major role in enhancing atomization in the combustion chamber, which can be considered as an effective way of promoting efficient combustion. The engine experiments have confirmed the overall benefits of micro-explosion of emulsion fuels [13-15]. Several numerical and experimental studies on emulsion fuel droplets were performed to understand the micro-explosion phenomenon [16-23]. Shinjo et al. [23] investigated the physics of puffing and micro-explosion of emulsion fuel droplets, and they reported that it might be possible to control micro-explosion/puffing in a fuel spray by the appropriate mixing of fuel blends and ambient flow conditions. An extensive investigation of micro-explosion phenomenon in multi-component miscible fuels has also been performed [24-32]. Lasheras et al. [24] showed that free droplets of alcohol/n-paraffin solutions and emulsions could undergo disruption phenomenon. They demonstrated that a minimum difference exists between the boiling points of the low volatile and high volatile fuel components along with a suitable range of relative concentration of alcohol and n-paraffin to achieve disruptive burning. Wang and Law [27] observed that micro-explosion is possible only if alcohol is the more volatile, lower boiling point component in the mixture of alkanes and alcohols. Their work indicated that using light alcohols like methanol and ethanol as additives could improve the atomization. Shen et al. [32] proposed a numerical model of micro-explosion in multi-component bio-fuel droplets. The proposed model was used to characterize the onset of micro-explosion by the normalized onset radius (NOR). They also estimated the Sauter mean radius (SMR) of the secondary droplets formed due to micro-explosion of primary or parent droplet.

Despite all the experimental studies on multi-component fuel droplets, little is known about the onset of nucleation, variation of bubble growth and diameter and velocity of ejected droplets from puffing and micro-explosion during the combustion of different blended fuels. In the present work, experimental investigations on the onset of nucleation, bubble growth and breakup of butanol/Jet A-1 and ABE/Jet A-1 blended droplets have been carried out in details. An attempt has been made to characterize the probable status of the physical events and a synergetic link amongst them for different compositions of the fuel blend.

## 2. Experimental Methodology

The experiments were conducted in a closed stainless steel chamber under quiescent atmospheric conditions. The chamber consists of two quartz windows for optical access. A schematic of the apparatus is shown in Fig. 1. A micro-pipette was used to generate constant volume droplets of 2.5 ± 0.05 µl with an equivalent diameter of 1.7 ± 0.1 mm. The droplet was suspended on a quartz fiber having 0.2 mm diameter. Because of the lower thermal conductivity of quartz (1.4 W/m K), it is assumed that the fiber does not work as a heat source to the fuel droplet in the later part of the burning process. The fuels used in this study are Jet A-1, butanol and A-B-E (acetone-butanol-ethanol), whose properties are shown in Table 1. Although Jet A-1 is a multi-component fuel, it was considered as a single component in the present work, in consideration of less significant volatility differential amongst the components.

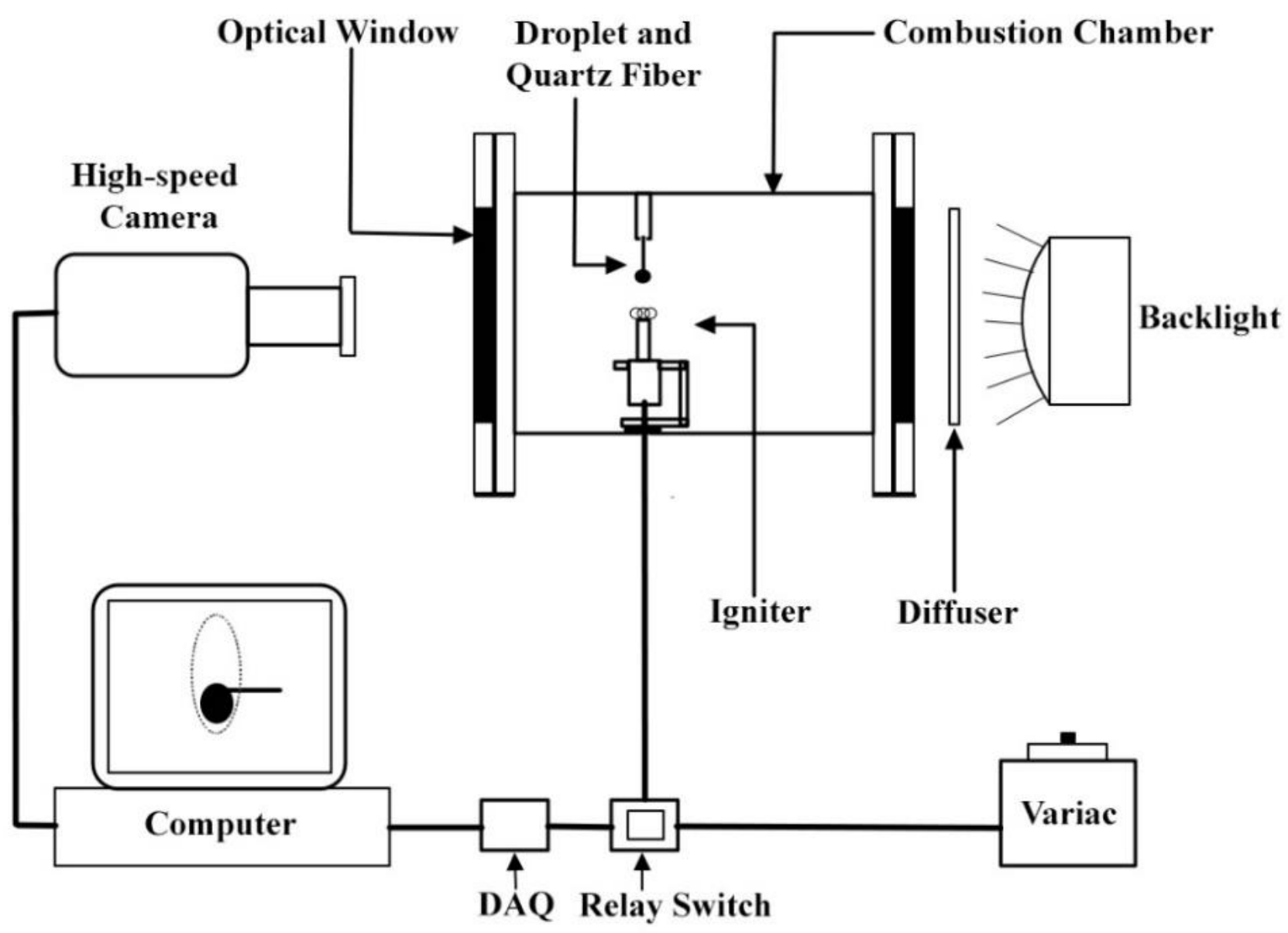

**Fig 1.** Schematic of the experimental apparatus.

Acetone, butanol, and ethanol were purchased commercially (Sigma-Aldrich) and A-B-E mixture was prepared at a ratio of 3:6:1 (A:B:E). Three blends of butanol/Jet A-1 and A-B-E/Jet A-1 were considered in the present study. The composition of fuel blends constituting the droplet is shown in Table 2. True color images were captured using a DSLR camera (at a frame rate of 25 fps) to differentiate the visual appearance of the flame for different blends. The burning sequence was captured using Phantom V7.3 high-speed monochrome camera at a resolution of 800 x 600 pixels at 3000 fps, and a Multi LED backlight is used to illuminate the droplets. An image analysis program, Image-Pro Plus, version 6.0 was used to determine the evolution of bubble diameter, and the diameter and velocity of ejected droplets. The uncertainty in the measurement of bubble diameter is ± 0.05 mm. This uncertainty arises mainly due to the non-spherical or asymmetric shape of the bubble. The uncertainty in the measurement of the secondary droplet diameter is ±10 µm. To guarantee the accuracy of the measurement and to verify the repeatability, the experiments were conducted 25 times for each blend case.

**Table 1**
Properties of the fuels investigated in this study.

| Physical Properties | Jet A-1[a] (Standard) | Acetone[b] | Butanol[c] | Ethanol[c] |
|---|---|---|---|---|
| Molecular Formula | $C_8$-$C_{16}$ | $C_3H_6O$ | $C_4H_{10}O$ | $C_2H_5OH$ |
| Boiling point (°C) | 180-250 | 56.1 | 117.7 | 78.4 |
| Density at15 °C (kg/m$^3$) | 775–840 | 791 | 813 | 795 |

[a]Standard specifications of Jet A-1 from ASTM D1655, [b]Properties of acetone are from Ref. [33,34] [c]Properties of ethanol and butanol are from Ref. [33,35].

**Table 2**
Composition and nomenclature of fuel blends

| Composition of fuel blends (volume basis) | Designated nomenclature |
|---|---|
| 10% butanol, 90% Jet A-1 | B10 |
| 30% butanol, 70% Jet A-1 | B30 |
| 50% butanol, 50% Jet A-1 | B50 |
| 30% acetone, 60% butanol, 10% ethanol | A-B-E |
| 10% A-B-E, 90% Jet A-1 | ABE10 |
| 30% A-B-E, 70% Jet A-1 | ABE30 |
| 50% A-B-E, 50% Jet A-1 | ABE50 |

# 3. Results and Discussion

## *3.1 Visual Appearance of the flame*

The sequence of flame images of fuel droplet comprising pure Jet A-1, butanol, A-B-E, B50, and ABE50 are shown in Fig. 2. The images show a typical envelope flame surrounding the droplet for all the cases. The flame of pure Jet A-1 appears to be relatively brighter yellowish with orange hue at the top edge, which is due to the emission from soot. However, the flame images of pure butanol and A-B-E show yellowish flame at the top edge and blue flame at the bottom. The appearance of the flames in case of B50 and ABE50 is somewhat different (less yellowish and absence of orange hue) than that of pure Jet A-1 case. This could be due to the fact that the combustion of B50 and ABE50 droplets produce lower soot compared to pure Jet A-1. The reduction in sooting propensity of butanol blends and ABE blends is attributed to the presence of oxygen atom in butanol molecule and constituent molecules of A-B-E. The sooting propensity is in the order as: Jet A-1 > ABE50 > B50 > A-B-E > pure butanol. As observed in Fig. 2 (i), the flame images indicate smooth burning of the jet fuel, although, jet fuel is a multi-component fuel consisting of several components with varying boiling points. It may be conjectured that the volatility differential between the components is not sufficient enough to initiate any nucleation. A-B-E and the blends of butanol/Jet A-1 and ABE/Jet A-1 show disruptive nature of burning due to the higher volatility differentials as evident from the sequence of flame images (Figs. 2(iii), 2(iv), and 2(v)).

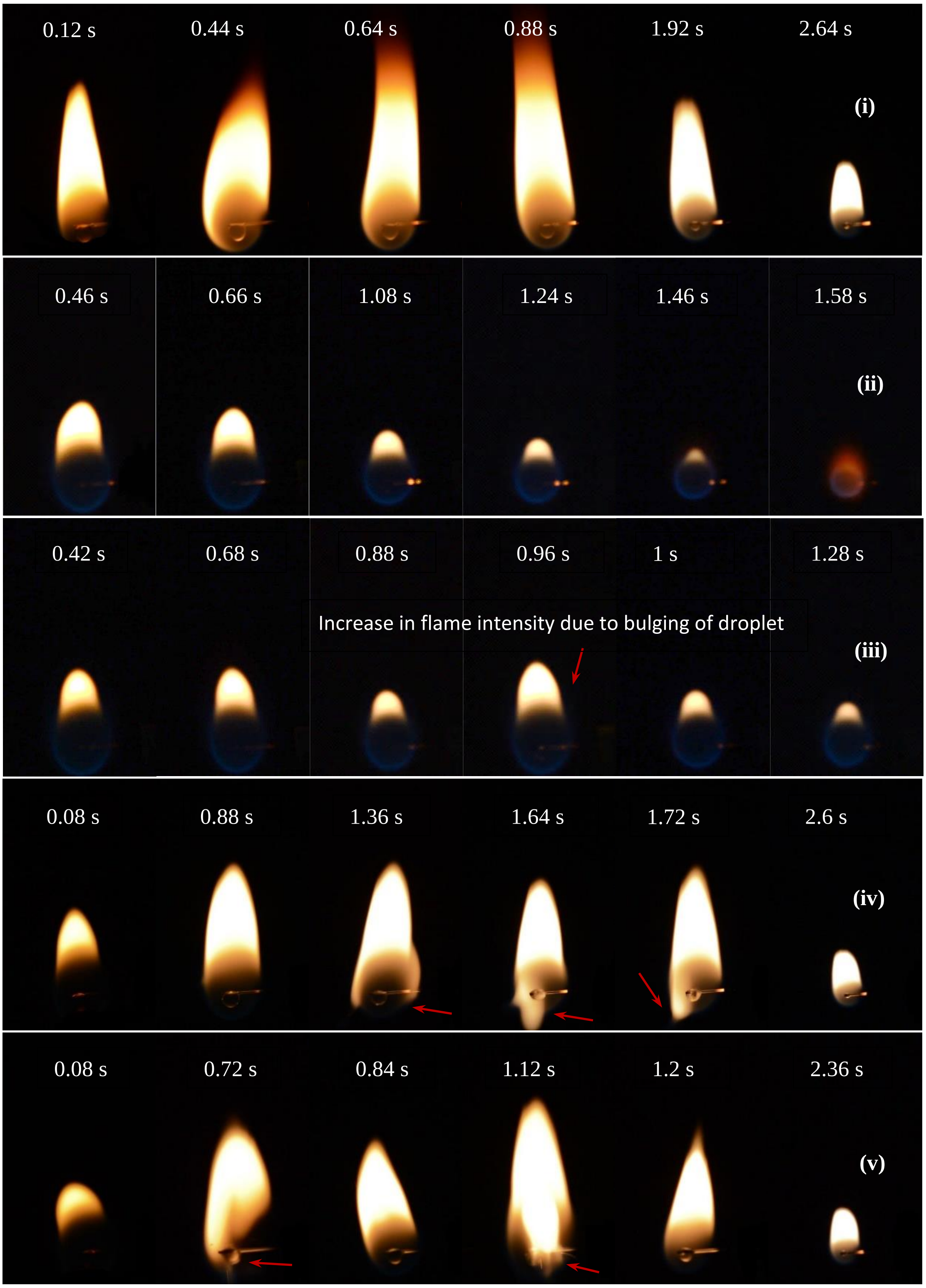

**Fig. 2.** The sequence of flame images indicating (a) smooth burning of (i) Pure Jet A-1, (ii) pure butanol and (b) disruptive burning of (iii) A-B-E, (iv) B50 and (v) ABE50 droplets. Arrows indicate disruptive behavior due to higher volatility differential.

## *3.2 Normalized Squared Onset Diameter*

Nucleation sites were observed mostly near the core of the fuel droplets. The onset of nucleation was characterized by normalized squared onset diameter (NOD), which is the square of the ratio of droplet diameter at nucleation to the initial droplet diameter. The NOD value was determined by carefully observing the first appearance of the bubble visible from the high-speed images. It was noted that for B10 and ABE10 fuel droplets, nucleation sites were not sustainable by the subsequent growth of bubbles and their coalescence. Figure 3 represents the probability of NOD for different blends. The most probable NOD values for B30, B50, ABE30 and ABE50 are almost same, which are 0.81, 0.78, 0.81 and 0.82 respectively. This implies that earlier nucleation is the most probable one for these blends; however, delayed nucleation in above blends have also been observed, which are less probable in nature.

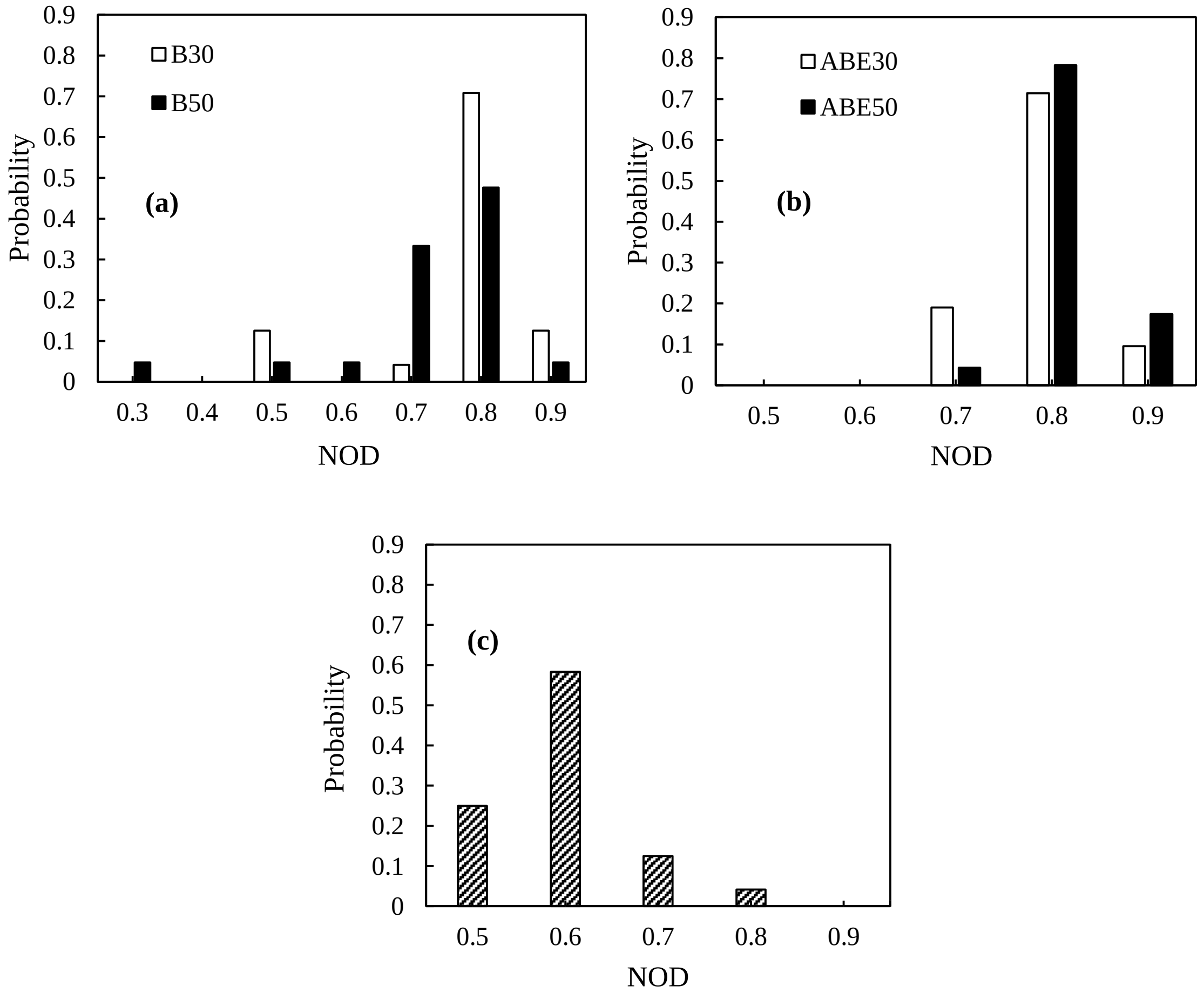

**Fig. 3.** The probability of NOD values for a) blends of butanol and b) blends of A-B-E and c) A-B-E.

The lower NOD value corresponds to the slow and higher degree of superheating of higher volatile component trapped inside the droplet, whereas the higher NOD value corresponds to relatively rapid and lower degree of superheating of the higher volatile component. The NOD value for A-B-E droplets is significantly lower than its blends with Jet A-1 since the volatility differential in A-B-E mixture i.e. between less volatile component (butanol), and more volatile component (acetone) is lower than the volatility differential in ABE blends with Jet A-1. The most probable NOD value for A-B-E is 0.6. All the results related to droplet regression, bubble growth leading to puffing/micro-explosion reported and discussed hereafter correspond to the most probable state of nucleation for a given fuel droplet.

### *3.3 Sequence of droplet burning, bubble growth, and evolution of droplet diameters*

The puffing and micro-explosion phenomena in different blends are represented by photographic sequences and evolution of droplet diameters. Figure 4 (a) and 4 (b) illustrates a schematic diagram of puffing and micro-explosion. Initially, the pressure inside the vapor bubble is high compared to the ambient liquid pressure. The tiny bubbles coalesce with relatively larger bubble due to the internal circulation inside the droplet. This leads to the formation of a bigger bubble. Since the nucleation sites are dependent on the proportion of higher volatile component, a lower proportion of the higher volatile component results in fewer nucleation sites and hence less coalescence of tiny bubbles.

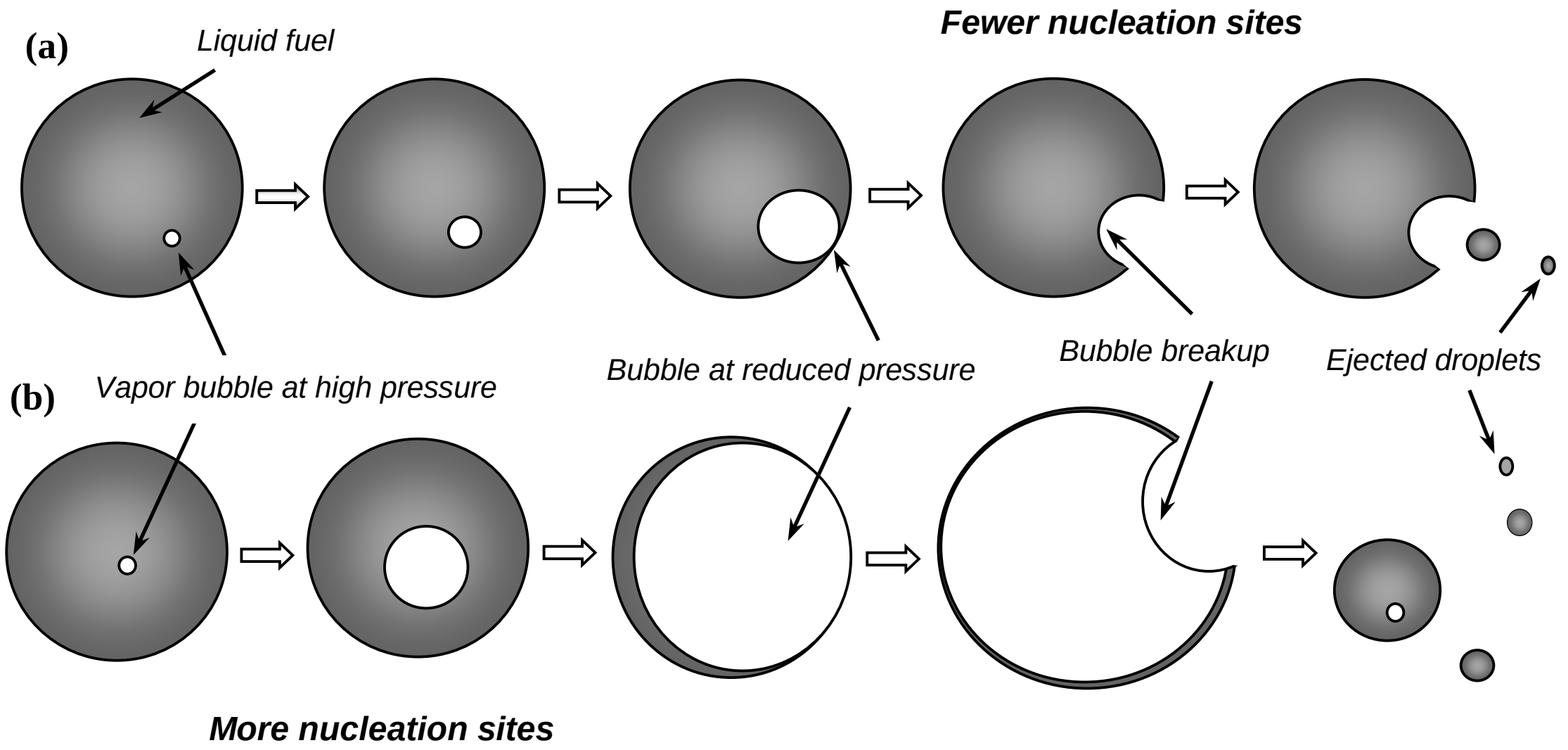

**Fig. 4.** Schematic diagram of (a) puffing and (b) micro-explosion. The sequence represents nucleation, bubble growth and breakup of parent droplet.

Therefore, the bubble cannot grow further and breaks apart resulting in the ejection of small secondary droplets. This breakup of a relatively smaller bubble is referred to puffing. A higher proportion of more volatile component causes a significant number of nucleation sites associated with coalescence of tiny bubbles. This results in the formation of a bigger bubble and its subsequent breakup leads to micro-explosion. In the present work, the photographic sequences were focused on the bubble growth and its breakup. Out of nearly 8000 frames, only a few frames have been selected and presented here. Figure 5 represents a typical sequence of images of bubble growth and micro-explosion of a typical A-B-E droplet, where $\tau_1$ = 0 ms represents the time of appearance of first bubble. The vapor bubble, which has grown from the nucleus, can be seen to be located nearly symmetrical within the droplet at $\tau_1$ = 13.6 ms. The bright spot at the center of the droplet is due to the significant difference in refractive index of the vapor and liquid and the subsequent scattering of light rays. A similar scattering of light highlights the bubble periphery which appears like a bright ring. The bubble can be observed to grow until it breaks at 50.6 ms, resulting in the ejection of secondary droplets at 52.3 ms. The nucleation occurs again, and as the bubble grows, a portion of its relatively smooth surface can be seen bulging out of the drop from 137-169.3 ms (Fig. 5).

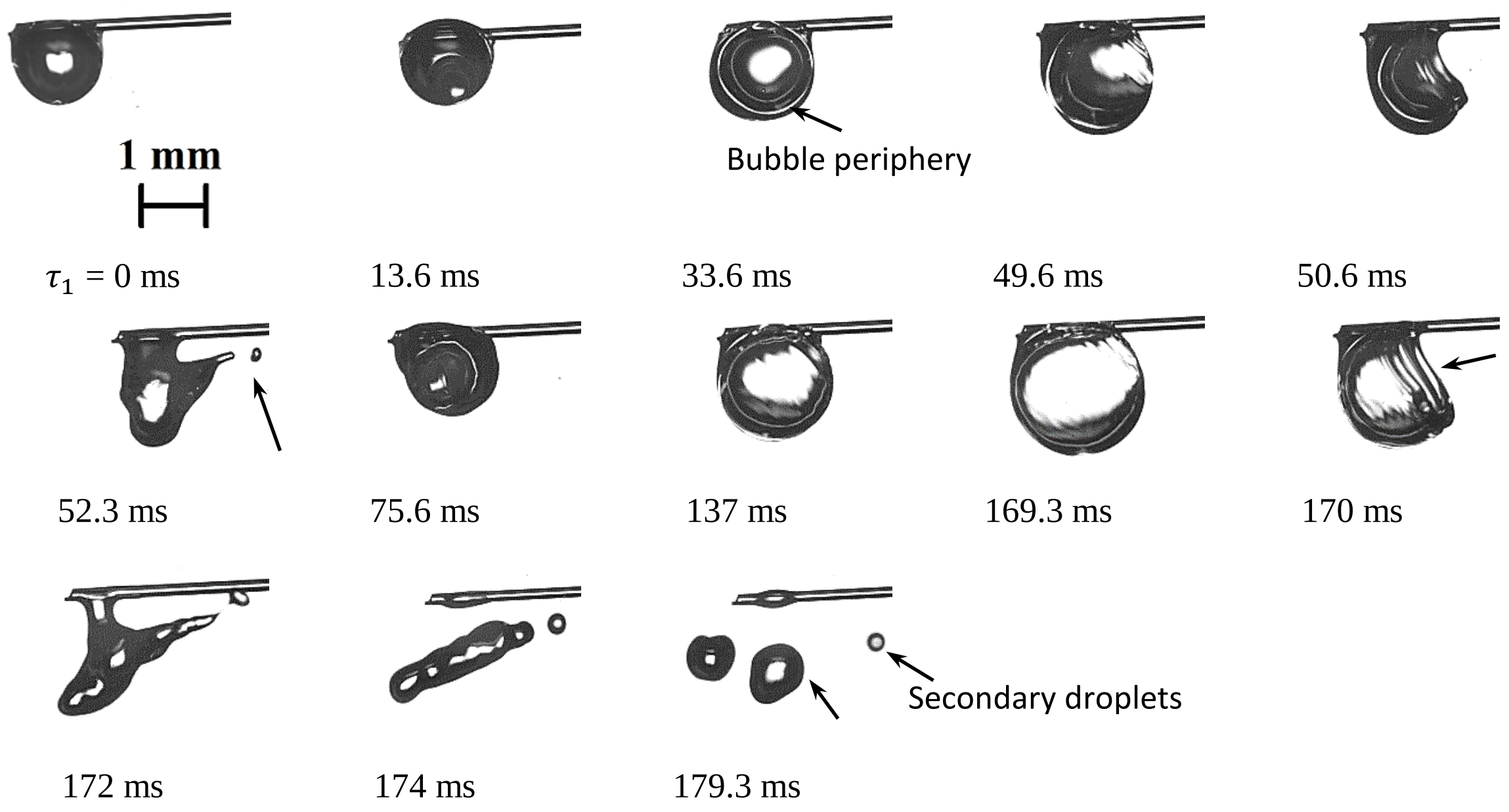

**Fig. 5.** The sequence of images of bubble growth and micro-explosion of a typical A-B-E droplet.

It is evident from the high-speed images that the occurrence of puffing has created turbulence inside the droplet resulting in the formation of more nucleation sites and in turn leading to the growth of a bigger bubble. A similar observation has been made by other researchers [23,36]. As the bubble grows and droplet continues to evaporate, the peripheral liquid sheet becomes thinner and at the same time is subjected to an internal pressure higher than that of the ambiance. This causes the tearing of the liquid sheet into small droplets at a later stage (at 179.3 ms in Fig. 5).

Figure 6 (a) represents a typical sequence of the combustion of B30 droplets indicating continuous puffing. It can be seen that the vapor is blown out of the droplet at different time intervals over most of the droplet lifetime. Although the volatility differential in B30 blend is sufficient for nucleation to occur, the proportion of butanol is not adequate for the bubble to grow significantly resulting in micro-explosion phenomenon. As the percentage of butanol is increased to 50%, the nucleation sites inside the droplet increase. A Higher number of nucleation sites favors the bubble growth and leads to micro-explosion. As found from repeated experiments, the probability of micro-explosion of B50 droplets is around 56% while the state of continuous puffing corresponds to the probability of 44%. A typical sequence of bubble growth and break-up of B50 case is shown in Fig. 6 (b). Initially, the vapor bubble grows nearly symmetrical until the heavier liquid distorts the spherically symmetrical shape while the lighter vapor moves upwards inside the droplet. It is important to note here that the nucleation in this situation occurs earlier in droplet's lifetime at NOD of 0.8. The droplet breaks at about 60 ms after the nucleation as seen in Fig. 6 (b). Similar behavior was observed in the bubble growth of ABE30 and ABE50 droplets (Fig. 6 (c) and 6 (d)). The probability of micro-explosion is relatively higher (more than 90%) for ABE50 droplets as compared to B50 droplets. The probability of micro-explosion in ABE30 case is somewhat similar to that of B50 case. This can be attributed to the higher volatility differential among the components in ABE blends. The probability of micro-explosion in ABE30 blend is little over 50%. Interestingly, more than 90% runs of the ABE50 blend resulted in micro-explosion. The effect of higher volatility differential and a larger proportion of more volatile component seem to play a major role here. It is observed from Fig. 6 that an increase in the proportion of the more volatile component in fuel blend increases the bubble diameter and, in turn, the probability and intensity of droplet breakup.

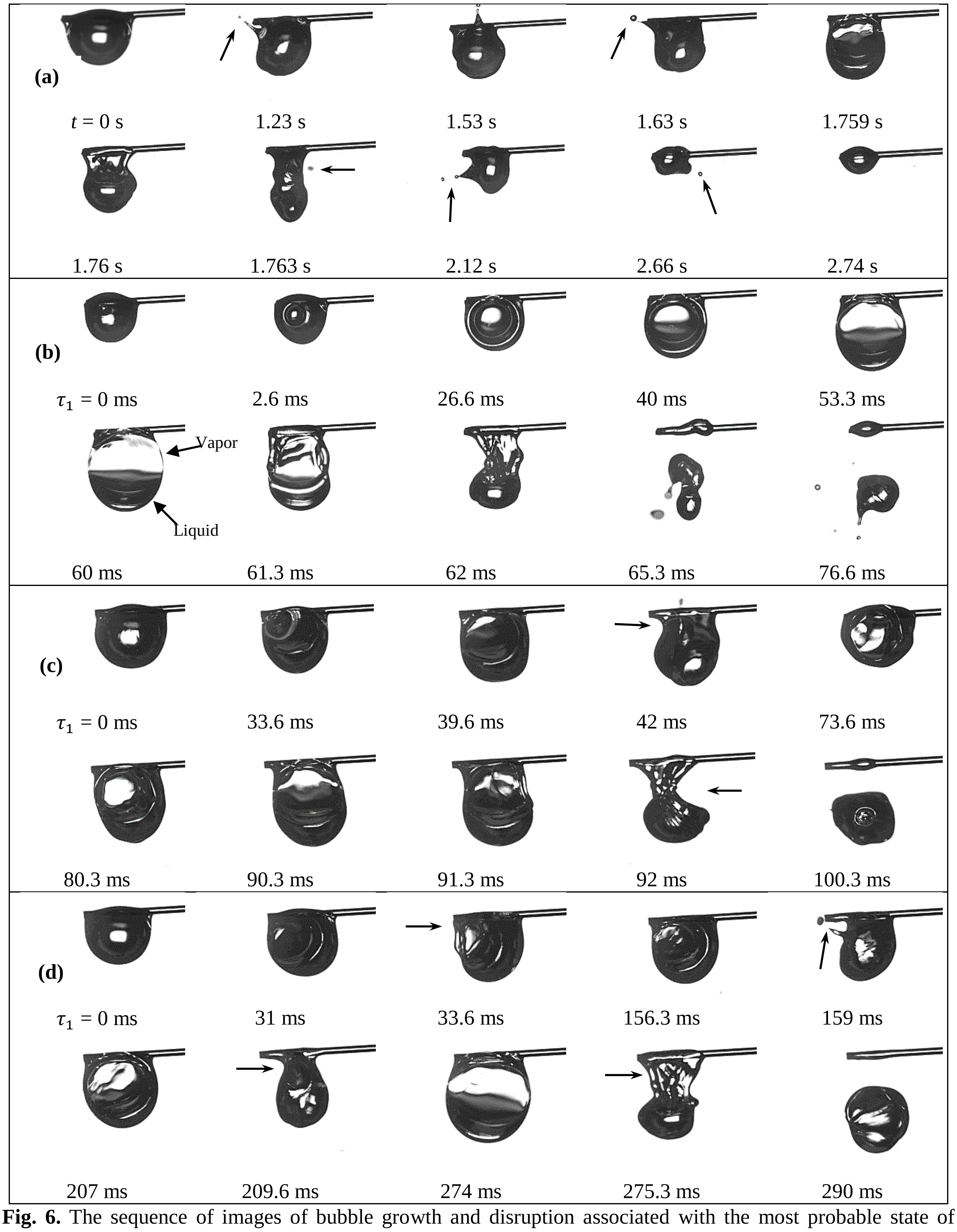

**Fig. 6.** The sequence of images of bubble growth and disruption associated with the most probable state of nucleation (a) B30, (b) B50, (c) ABE30; and (d) ABE50. The arrows indicate ejection of secondary droplets and breakup of parent droplet.

Bubble growth rate generally depends on surface tension, liquid inertia, and the pressure differential between the bubble and the surrounding liquid. After the early bubble growth, heat diffusion from the ambient liquid to the bubble becomes controlling factor [18,37]. The bubble growth for the fuel droplets corresponding to the most probable state of nucleation is shown in Fig. 7. From the bubble growth curves, two phases of growth (inertial growth and diffusion control growth) can be identified. The curves shown in Fig. 7 correspond to the sequence of images shown in Fig. 6 (a)-(d). It is evident that the bubble growth rates for B30, B50, ABE30, and ABE50 fuel droplets are almost the same; however, the growth rate is slow in the case of A-B-E. This slower growth rate might be due to the lower volatility differential among the components of A-B-E compared to that of butanol and A-B-E blends. It is observed that the final bubble diameter (prior to droplet breakup) is not same for these fuel blends despite having an almost identical bubble growth rate pattern. The final bubble diameter for B30, B50, ABE30, ABE50, and A-B-E are 1.70 mm, 2.16 mm, 1.92 mm, 2.19 mm, and 2.06 mm respectively.

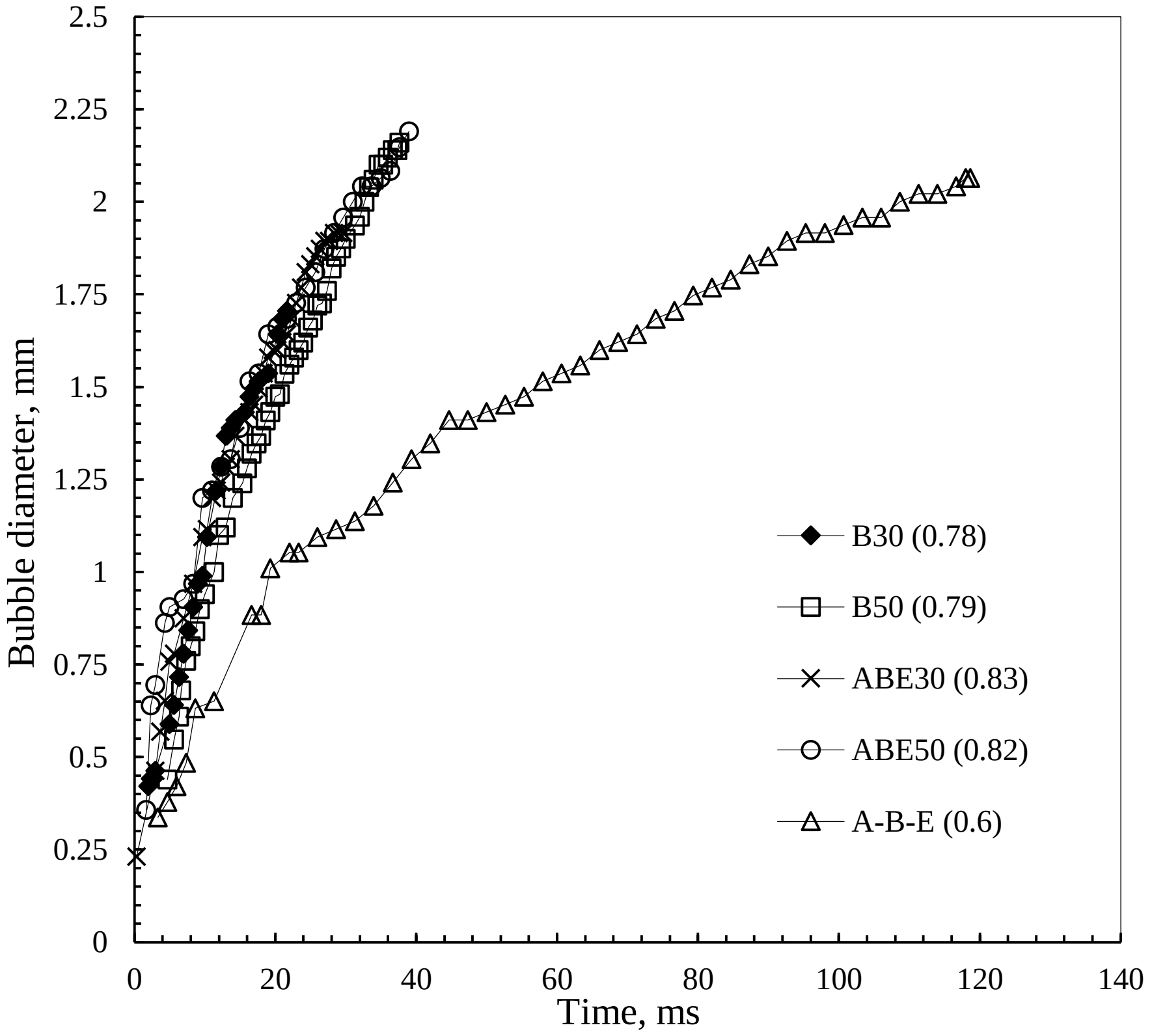

**Fig. 7.** Variation of bubble diameter with time for different blends.

Figure 8 shows a comparison of the evolution of droplet diameters for pure Jet A-1, pure butanol, and A-B-E. The droplets of Jet A-1 and pure butanol show a smooth and continuous temporal regression of diameter due to evaporation without disruption by bubble formation. It is observed that the droplet lifetime of pure butanol is nearly equal to that of pure Jet A-1. The similar lifetime can probably be attributed to the fact that higher vapor pressure of butanol is being counterweighted by its higher enthalpy of vaporization compared to that of Jet A-1.

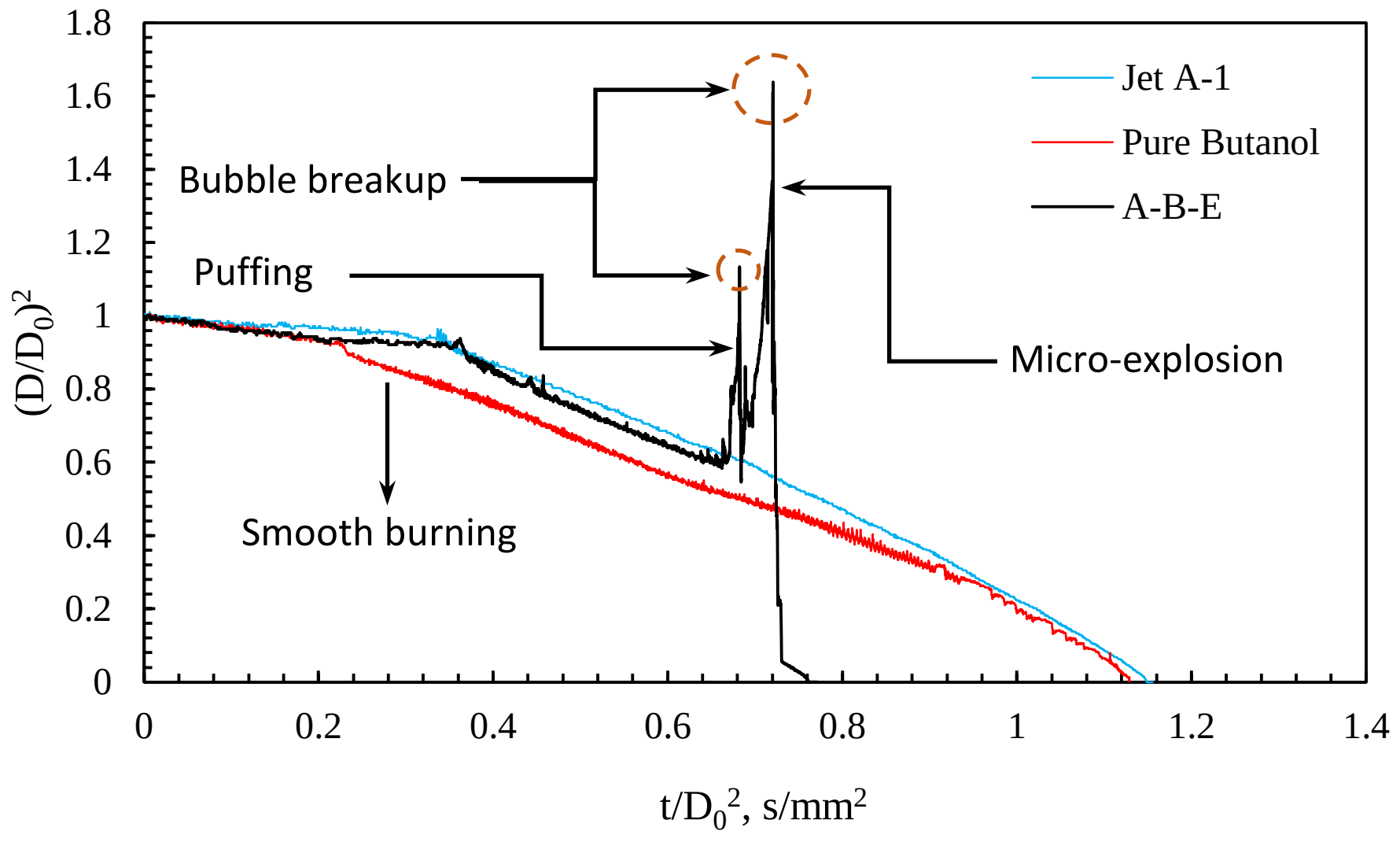

**Fig. 8.** Comparison between the temporal evolutions of droplet diameter of typical Jet A-1, pure butanol, and A-B-E droplets.

On the other hand, two prominent spikes as found in Fig. 8, characterize the regression rate of A-B-E. The sudden increase in droplet diameter followed by its immediate rapid decrease is due to the expansion and subsequent breakup of the bubble. The first spike corresponds to the bubble growth leading to puffing phenomenon while the second spike corresponds to the micro-explosion phenomenon. The spikes represent the maximum droplet diameter at which the bubble breaks apart. It is noticeable that the time from bubble generation to the final breakup is very short (of the order of 1/100 of the average droplet lifetime). The drop size after the breakup is significantly smaller, and thus, the vaporization is greatly enhanced. The droplet lifetime of A-B-E is around 0.64 times to that of pure Jet A-1 droplet. Figure 9 represents the temporal evolution of the droplet diameter of butanol blends and A-B-E blends that correspond to the sequence of images shown in Fig. 6. The continuous fluctuations in Fig. 9 (a) represents continuous puffing, which is caused by the ejection of secondary droplets. As seen in Fig. 9 (b) to 9 (d), whenever micro-explosion occurs, puffing precedes it. As discussed before, puffing enhances the turbulent mixing inside the droplet. This mixing helps in creating more nucleation sites, which in turn leads to the formation of a bubble sufficient for micro-explosion.

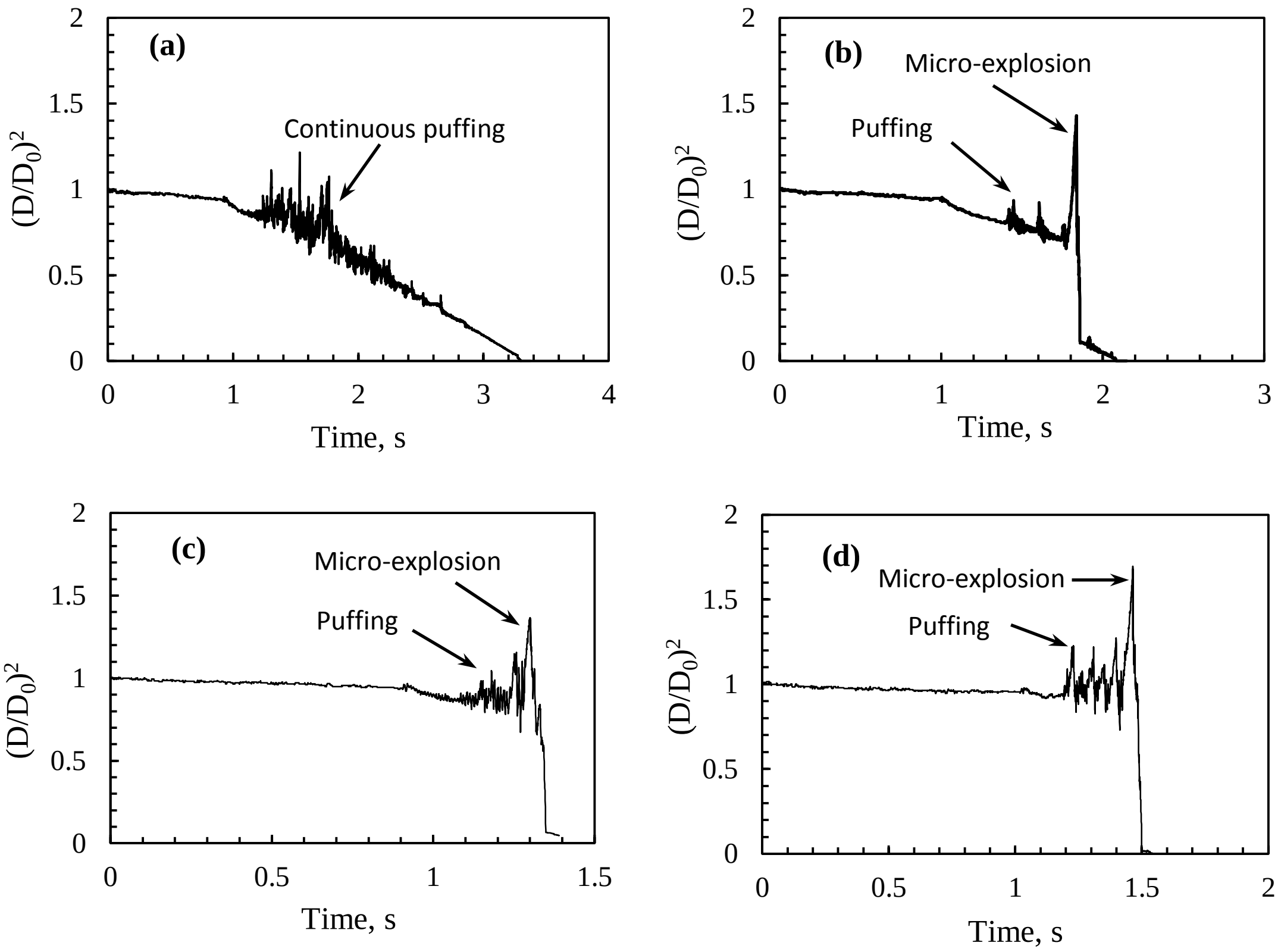

**Fig. 9.** The temporal evolution of droplet diameter for a) B30, b) B50, and c) ABE30, and d) ABE50 compositions associated with most probable NOD.

The characteristic features of disruptive droplet burning comprising different fuel blends are highlighted in Table 2. It is noticeable that the micro-explosion leads to a significant reduction in droplet lifetime. A trend is observed from Table 2 that the increase in the proportion of the higher volatile component in the blends increases the probability of micro-explosion and hence decreases average droplet lifetime.

**Table 2**

Characteristics features of disruptive burning in different blends

| Blends | Dominant characteristics of burning | Probability of Micro-explosion | Avg. droplet lifetime (relative to pure Jet A-1) |
|---|---|---|---|
| B30 | Continuous puffing | 12% | 92% |
| B50 | Micro-explosion | 56% | 55% |
| ABE30 | Micro-explosion | 52% | 72% |
| ABE50 | Micro-explosion | 92% | 49% |
| A-B-E | Micro-explosion | 64% | 64% |

*3.4 Diameter and velocity of ejected droplets*

The diameter versus velocity distribution of ejected droplets during puffing and micro-explosion is shown in Fig. 10. It was observed that the puffing was the dominant phenomenon for B30 droplets while micro-explosion was dominant for other fuel droplets.

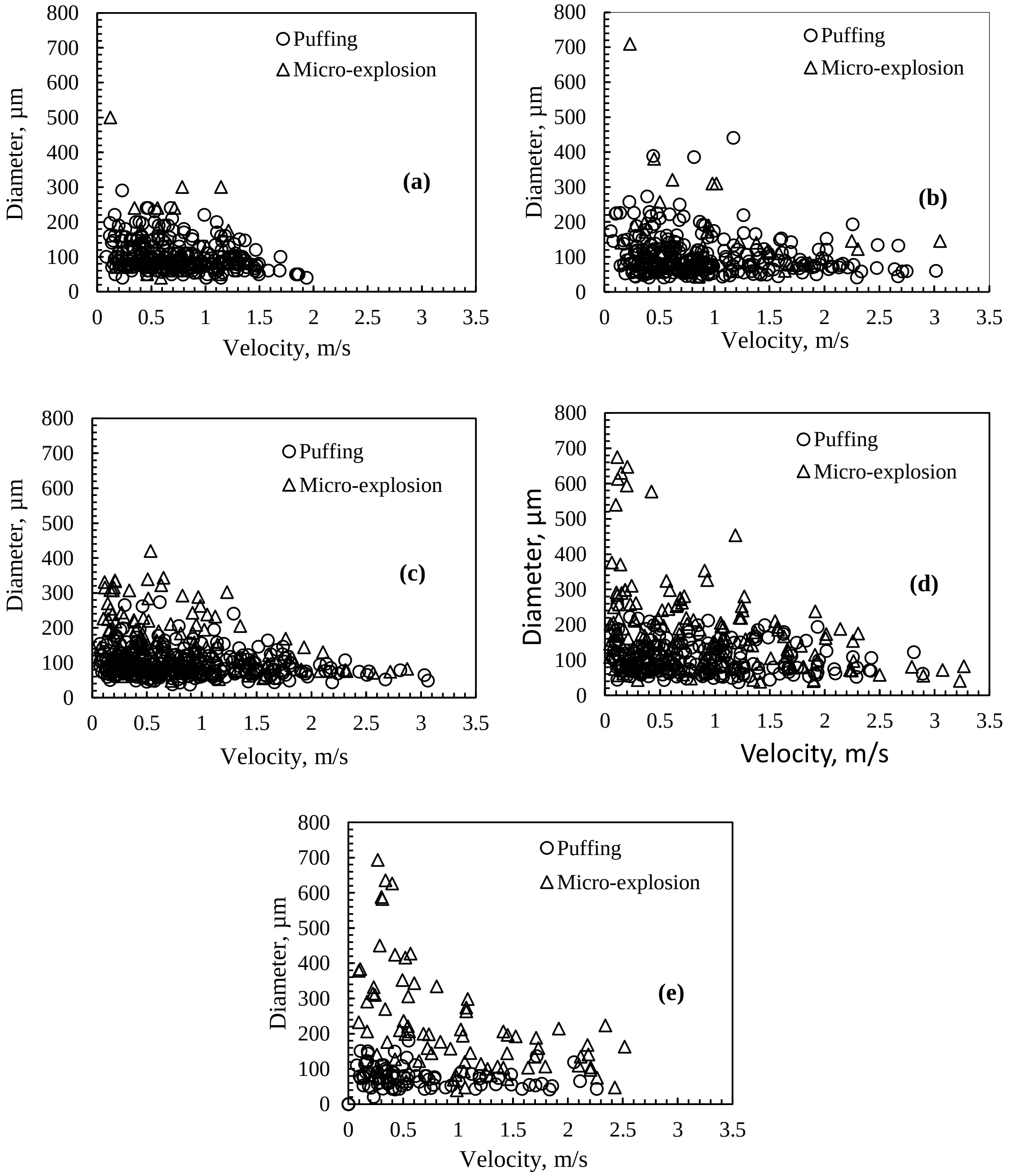

**Fig. 11.** Diameter vs velocity distribution of ejected droplets due to the puffing and micro-explosions in a) B30, b) B50, c) ABE30, d) ABE50, and e) A-B-E droplets.

However, micro-explosion was always preceded by puffing. It is evident from Fig. 10 that both puffing and micro-explosion resulted in the ejection of multiple droplets with both larger and smaller diameters; however, puffing seems to produce relatively smaller diameter droplets compared to micro-explosion. The Sauter mean diameters (SMD) of ejected droplets are 170, 290, 130, 290, and 380 μm for B30, B50, ABE30, ABE50, and A-B-E respectively.

### *3.5 Puffing/Micro-explosions in secondary droplets*

Multiple puffing/micro-explosions were observed in 30% (B30 and ABE30) and 50% (B50 and ABE50) blends. Secondary micro-explosions has been reported previously by researchers [38] during the evaporation of multicomponent droplets at high ambient temperature; however, the bubble growth was not indicated in the secondary droplets. The secondary puffing/micro-explosion reported by the researchers is rather abrupt. In the present work, bubble growth was observed in the secondary droplets which led to the subsequent breakup of the secondary droplet. The secondary droplets created from the first micro-explosion of the parent droplet sometimes remained out of frame. Some selected sequences of images of multiple puffing/micro-explosions in B50, ABE30, and ABE50 blends are shown in Fig. 11 (a)-(c). Here, $\tau_2$ = 0 ms represents the onset of micro-explosion in parent droplet. As seen in Fig. 11, after the occurrence of first puffing or micro-explosion, a bubble starts to grow in the separated droplet and subsequently breaks apart leading to the ejection of secondary droplets. Due to the presence of a substantial proportion of volatile component in the secondary droplets, the bubble forms and starts to grow again in those droplets leading to second micro-explosion.

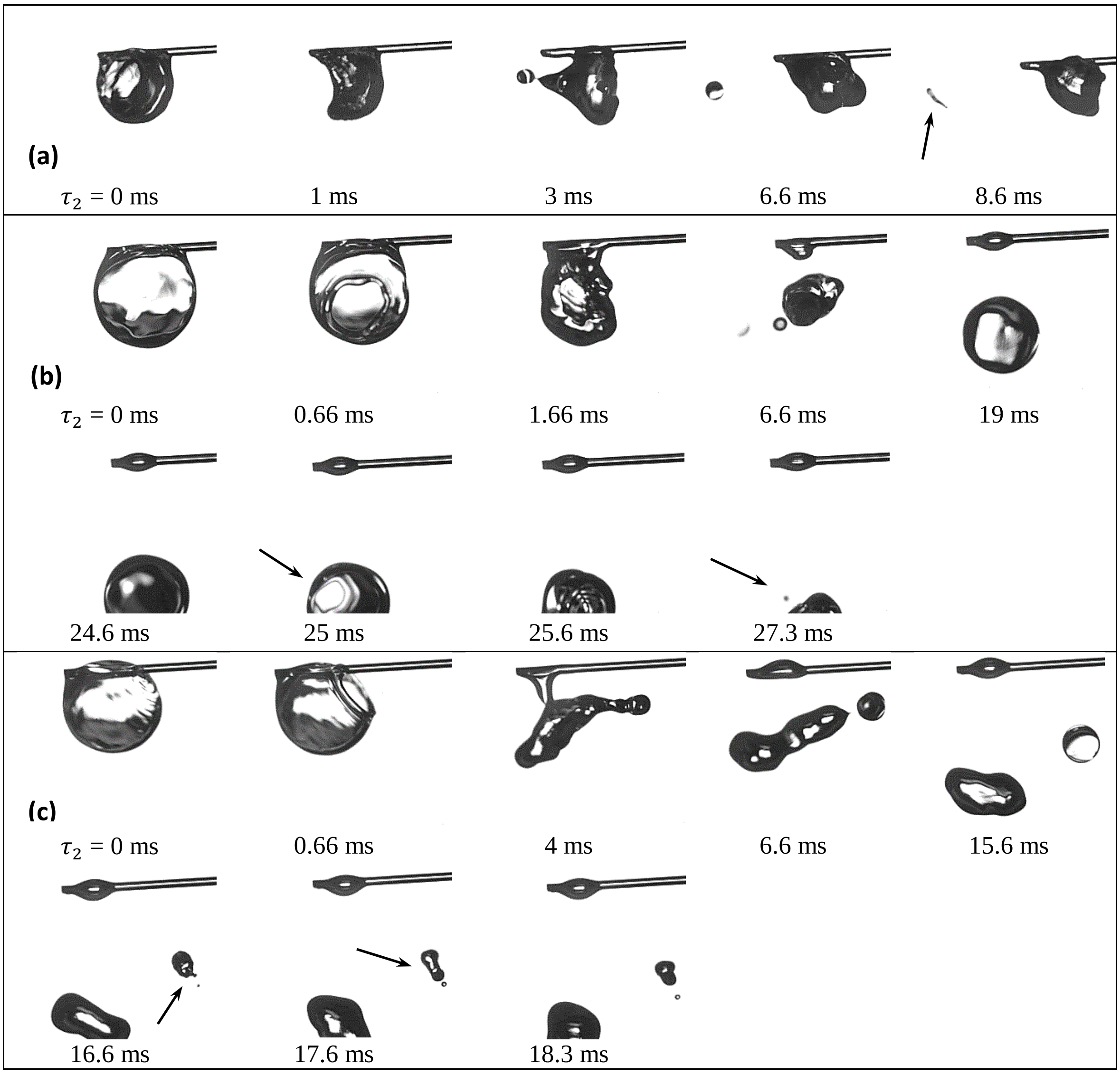

**Fig. 11.** The sequence of images of puffing/micro-explosions in secondary droplets of (a) B50, (b) ABE30, and (c) ABE50 droplets.

# 5. Conclusions

An experimental investigation has been carried out on droplet combustion of butanol/Jet A-1 and A-B-E/Jet A-1 blends. The major observations relating to different states of droplet in course of its combustion are as follows:

(1) Smooth burning was observed in droplets comprising Jet A-1, pure butanol, B10, and ABE10, whereas disruptive burning was observed in B30, B50, ABE30, ABE50, and A-B-E droplets. The nucleation in A-B-E was observed to be delayed as compared to that in butanol/Jet A-1 and A-B-E/Jet A-1 blends.

(2) Puffing was observed to be the dominant phenomenon for B30 while micro-explosion was the dominant phenomenon in droplets with other blends (B50, ABE30, and ABE50).

(3) The bubble growth rate is almost the same for B30, B50, ABE30 and ABE50 droplets. For the droplets with 30% blends (B30 and ABE30), the period of bubble growth is smaller and hence the final bubble diameter is smaller. For droplets with 50% blends (B50 and ABE50), the bubble growth period is longer resulting in higher final bubble diameter.

(4) The probability of occurrence of micro-explosion in A-B-E blends was observed to be greater than that of butanol blends.

(5) Puffing seemed to produce secondary droplets of smaller size as compared to those in micro-explosion.

(6) Puffing/micro-explosions phenomena were observed in the secondary droplets as well.

## Acknowledgements

The authors would like to acknowledge Indian Institute of Technology Kharagpur for providing financial support through Innovative Research and Development (ISIRD) programme.